\documentclass{PoS}
\pdfoutput=1
\newcommand{\cd}{\makebox[0.08cm]{$\cdot$}}
\newcommand{\sla}{\not\!}
\title{Chiral effective field theory on the light front}

\ShortTitle{Chiral effective field theory on the light front}

\author{\speaker{J.-F. Mathiot}\\
Laboratoire de Physique Corpusculaire, Universit\'e Blaise Pascal,
63177 Aubi\`ere Cedex, France\\
E-mail: \email{mathiot@in2p3.fr}}

\abstract{We propose a new approach to describe baryonic structure in terms of an effective chiral Lagrangian. The state vector of a baryon is defined on the light front of general position $\omega \cd x=0$, where $\omega$ is an arbitrary light-like four vector. It is then decomposed in Fock components including an increasing number of pions. The maximal number of particles in the state vector is mapped out to the order of decomposition of the chiral effective Lagrangian to have a consistent calculation of both the state vector and the effective Lagrangian. An adequate Fock sector dependent renormalization scheme is used in order to restrict all contributions within the truncated Fock space. To illustrate our formalism, we calculate the anomalous magnetic moment of a fermion in the Yukawa model in the three-body truncation. We present perspectives opened by the use of a new regularization scheme based on the properties of fields as distributions acting on specific test functions.}

\FullConference{International Workshop on Effective Field Theories: from the pion to the upsilon \\
		February 2-6 2009\\
		Valencia, Spain}

\begin{document}

\section{Introduction}
The calculation of baryon properties within the framework of chiral perturbation theory is the subject of active theoretical developments. Since the nucleon mass is not zero in the chiral limit, all momentum scales are involved in the calculation of baryon properties (like masses or electro-weak observables) beyond tree level. This is at variance with the meson sector for which a meaningfull power expansion of any physical amplitude can be done.
While there is not much freedom, thanks to chiral symmetry, for the construction of the effective Lagrangian in Chiral Perturbation Theory (CPT), ${\cal L}_{CPT}$, in terms of the pion field - or more precisely in terms of the U field defined by $U=e^{i {\bf \tau}.{\bf \pi}/f_\pi}$ where $f_\pi$ is the pion decay constant - one should settle an appropriate approximation scheme in order to calculate baryon properties. Up to now, two main strategies have been adopted. The first one is to force the bare (and hence the physical) nucleon mass to be infinite, in Heavy Baryon Chiral Perturbation Theory \cite {manohar}. In this case, by construction, an expansion in characteristic momenta can de developped. The second one is to use a specific regularization scheme \cite{IR} in order to separate contributions which exhibit a meaningfull power expansion, and hide the other parts in appropriate counterterms.
In both cases howether, the explicit calculation of baryon properties relies on an extra approximation in the sense that physical amplitudes are further calculated by expanding ${\cal L}_{CPT}$ in a finite number of pion field. 

Moreover, it has recently been realized that the contribution of pion-nucleon resonances, like the $\Delta$ and Roper resonances, may play an important role in the understanding of the nucleon properties at low energies \cite{delta}. These resonances are just added "by hand" in the chiral effective Lagrangian. This is also the case for the most important $2\pi$ resonances, like the $\sigma$ and $\rho$ resonances. 

\begin{figure}[btph] 
\begin{center} 
\includegraphics[width=18pc]{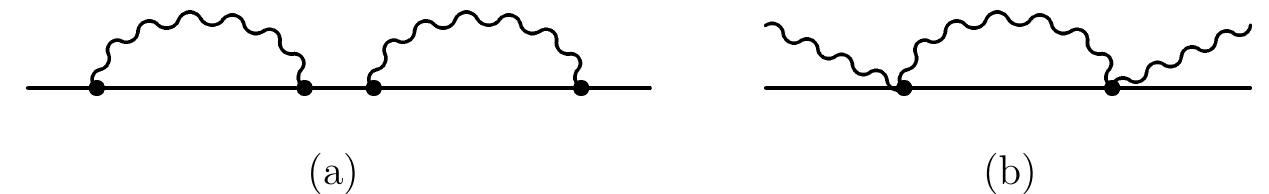}
\caption{Iteration of the self-energy contribution in first order perturbation theory (a); irreducible contribution to the bound state equation (b).\label{self_gen}}
\end{center}
\end{figure}

We would like  to propose in the following a new formulation in order to describe baryon properties in a systematic way. Since in the chiral limit the pion mass is zero, any calculation of $\pi N$ systems demands a relativistic framework to get, for instance, the  right analytical properties of the physical amplitudes. The calculation of bound state systems, like a physical nucleon composed of a bare nucleon coupled to many pions, relies also on a non-perturbative eigenvalue equation. While the mass of the system can be determined in leading order from the iteration of the $\pi N$ self-energy calculated in first order perturbation theory, as indicated in Fig.\ref{self_gen}.a, this is in general not possible in particular for $\pi N$ irreducible contributions  as shown on Fig.\ref{self_gen}.b.
%

\section{Light-front chiral effective field theory}
The general framework to deal with these requirements is Light Front Dynamics (LFD). 
Relativistic description admits some freedom in choosing
the space-like hyper-surface on which the state vector is
defined~\cite{dirac}.  In the
standard version of LFD, the state vector is defined on the plane
$t+\displaystyle{z}= 0$,
invariant with respect to Lorentz boosts along the $z$ axis.
Since the
vacuum state in LFD coincides with the free vacuum, one can
construct any physical system in terms of combinations of
free fields, i.e. the state vector is decomposed in a series of
Fock sectors with an increasing number of constituents. Note that the triviality of the vacuum in LFD does
not prevent from non-perturbative zero-mode contributions  to field operators \cite{hksw}. 

\begin{figure}[btph] 
\begin{center} 
\includegraphics[width=18pc]{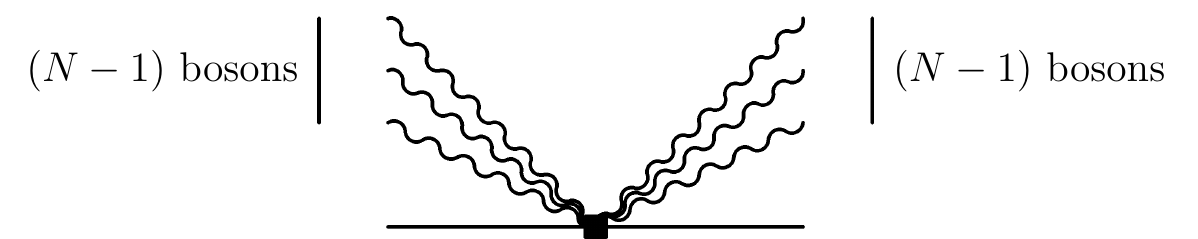}
\caption{General vertex including a maximum of $(N-1)$ pion fields in the initial and final states. \label{counting}}
\end{center}
\end{figure}

This decomposition of the state vector in a finite number of Fock components implies to consider an effective Lagrangian which enables all possible elementary couplings between the pion and the nucleon fields. This is indeed easy to achieve in chiral perturbation theory since each derivative of the U field involves one derivative of the pion field. In the chiral limit, the chiral effective Lagrangian of order $p$ involves $p$ derivatives and at least $p$ pion fields. In order to calculate the state vector in the $N$-body truncation, with one fermion and $(N-1)$ pions, one has to include contributions up to $2(N-1)$ pion fields in the effective Lagrangian, as shown on Fig.~\ref{counting}. We thus should calculate the state vector in the $N$-body truncation with an effective Lagrangian, denoted by ${\cal L}_{eff}^N$, and given by
\begin{equation}
{\cal L}_{eff}^N={\cal L}_{CPT}^{p=2(N-1)}\ .
\end{equation}
For a non zero pion mass, one should extend this mapping by taking $m_\pi^2$ of order $p=2$, as it is done in CPT.

While the effective Lagrangian in Light-Front Chiral Effective Field Theory (LF$\chi$ EFT) can be mapped out to the CPT Lagrangian of order $p$, the calculation of the state vector does not rely on any momentum decomposition. It relies only on an expansion in the number of pions in flight at a given light-front time. In other words, it relies on an expansion  in the fluctuation time, $\tau_f$, of such contribution. From general arguments, the more particles we have at a given light-front time, the smaller the fluctuation time is. At low energies, when all processes have characteristic interaction times larger than $\tau_f$, this expansion should be meaningfull.

It is interesting to illustrate the general features of  LF$\chi$ EFT calculations. At order $N=2$, we already have to deal with irreducible contributions as shown on Fig.\ref{self_gen}.b. The calculation at order $N=3$ explicitly incorporates contributions coming from $\pi \pi$ interactions, as well as all low energy $\pi N$ resonances, like the $\Delta$ or Roper  resonances. Indeed, in the $\vert \pi \pi N>$ Fock sector, the $\pi N$ state can couple to both $J=T=3/2$ as well as $J=T=1/2$ states. We can generate therefore all $\pi N$ resonances in the intermediate state without the need to include them explicitly, provided the effective Lagrangian has the right dynamics to generate these resonances. This is the case, by construction,  in CPT.

To settle a general framework based on LFD, one has however to address three different problems: {\it i)} one has to control in a systematic way the violation of rotational invariance; {\it ii)} one needs an adequate renormalization scheme consistent with the truncation of the Fock space;
{\it iii)} one should use an appropriate systematic regularization scheme which preserves all symetries.
We shall adress these three problems in the following.

\section{Covariant formulation of light-front dynamics}
In the Covariant Formulation of Light-Front Dynamics (CLFD) \cite{cdkm}, the state vector is defined on the light-front plane of general
orientation  $\omega\cd x=0$, where $\omega$  is  an arbitrary
light-like four-vector $\omega^2$=0.
The state vector, $\phi_\omega^{J\sigma}(p)$,
of a bound system corresponds to definite values for the
mass $m$, the four-momentum $p$, and the total angular momentum
$J$ with projection $\sigma$ onto the $z$ axis in the rest frame,
i.e., the state vector forms a representation of the Poincar\'e
group. The four-dimensional angular momentum
operator,  $\hat{J}$, is represented as a sum of the free and interaction parts:
\begin{equation}
\label{kt2} \hat{J}_{\rho\nu}=\hat{J}^{(0)}_{\rho\nu}
+\hat{J}^{int}_{\rho\nu}\ .
\end{equation}
From the general transformation properties of both the  state
vector and the LF plane, we have:
\begin{equation}\label{kt12}
\hat{J}^{int}_{\rho\nu} \ \phi_\omega^{J\sigma}(p)=
\hat{L}_{\rho\nu}(\omega)\phi_\omega^{J\sigma}(p) \ ,
\end{equation}
where
\begin{equation}\label{kt13}
\hat{L}_{\rho\nu}(\omega) =i\left(\omega_{\rho}
\frac{\partial}{\partial\omega^{\nu}} -\omega_{\nu}
\frac{\partial}{\partial\omega^{\rho}}\right)\ .
\end{equation}
The equation~(\ref{kt12}) is called the {\it angular condition}.
This equation does not contain the interaction Hamiltonian,
once  $\phi$ satisfies the Poincar\'e group equations \cite{cdkm}. The construction of the wave functions of
states with definite total angular momentum becomes therefore {\it
a purely kinematical problem}. The dynamical dependence of the wave functions
on the light-front plane orientation now turns into their explicit
dependence on the four-vector $\omega$. Such a
separation, in a covariant way, of kinematical and dynamical
transformations is a definite advantage of CLFD as compared to
standard LFD on the plane $t+z=0$.
The eigenvalue equations for the Fock components can be obtained from
the Poincar\'e group equation \cite{kms_08}
\begin{equation} \label{P2M2}
\hat P^2 \phi(p) = m^2 \phi(p)\ .
\end{equation}
We now decompose the state vector of a physical
system in Fock sectors. Schematically, we have:
\begin{equation}
\phi(p) \equiv \vert 1 \rangle +
\vert 2 \rangle +
 \dots + \vert n \rangle + \dots
\end{equation}
Each term on the r.h.s. denotes a state with a fixed number of
particles. It is proportional to the Fock component, or many-body wave function, $\phi_n$.
The spin
structure of $\phi_{n}$ is very
simple, since it is purely kinematical, but it should incorporate
$\omega$-dependent components in order to fulfill the angular
condition~(\ref{kt12}).  In the Yukawa model  we have, for $N=2$:
\label{Yuphi}
\begin{eqnarray}
\label{oneone}
\Gamma_{1}^{(2)}&= &a_1\  \bar{u}(k_1)u(p)\ ,\\
\Gamma_{2}^{(2)}&=&\bar{u}(k_1) \left[b_1  +
b_2\ \frac{m \sla \omega }{\omega \cd p}\right] u(p)\ , \label{onetwo}
\end{eqnarray}
since no other independent spin structures can be constructed. In (\ref{oneone},\ref{onetwo}), we have used the vertex function $\Gamma_n^{(N)}=(s_n-M^2)\phi_n$ where $s_n=(k_1 +  \ldots k_n)^2$.
Here $u$'s are free bispinors of mass $m$, $a_1$, $b_1$, and $b_2$ are
scalar functions determined by the dynamics.

\section{Non-perturbative renormalization scheme in light-front dynamics}
\begin{figure}[btph]
\begin{center}
\includegraphics[width=20pc]{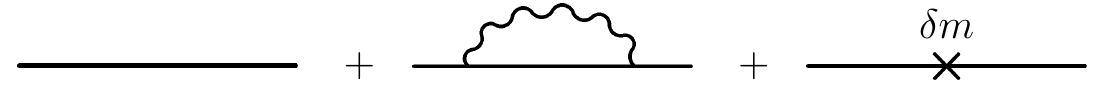}
\caption{Renormalization of the fermion propagator in second
order perturbation theory.\label{self}}
\end{center}
\end{figure}

In order to be able to make definite predictions for physical
observables, one should also define a proper renormalization
scheme. This should be done with care since the Fock decomposition
of the state vector is  truncated to a given order. Indeed,
looking at Fig.~\ref{self} for the calculation of the fermion
propagator in second order
perturbation theory, one immediately realizes that the
cancellation of divergences between the self-energy contribution
(of order two in the Fock decomposition) and the fermion Mass
Counterterm (MC) (of order one) involves two different Fock
sectors \cite{kms_08}.
This means that any MC and, more generally, any Bare Coupling
Constant (BCC) should be associated  with the number of particles
present (or ``in flight'') in a given Fock sector. In other words,
all MC's and BCC's must depend on the Fock sector under
consideration. The original MC $\delta m$ and the
fermion-boson BCC $g_0$ should thus be extended to a whole
series:
\begin{equation} \label{FSDR}
 \delta m  \to \delta m^{(i)} \ \ \ \  ,\ \ \ \  g_0 \to  g_0^{(i)}
\end{equation}
where $i=1,2,\ldots N$ 
refers  to the number of particles in ``flight''. A
calculation of order $N$ involves $\delta m^{(1)}\ldots \delta
m^{(N)}$ and $g_0^{(1)} \ldots g_0^{(N)}$. The quantities $\delta m^{(i)}$ and
 $g_0^{(i)}$ are calculated by
solving the systems of equations for the vertex functions in
the $N=1,2,3 \ldots$ approximations successively.
This procedure, which we call Fock Sector Dependent
Renormalization, is a well defined, systematic and
non-perturbative scheme \cite{kms_08}. 
The MC is determined from Eq.(\ref{P2M2}), while the
BCC $g_0^{(N)}$  is determined by demanding that the
$\omega$-independent part of the two-body vertex function $\Gamma_2$ at $s_2\equiv
(k_1+k_2)^2=m^2$
is proportional to the physical coupling constant $g$,
$ b_{1}(s_2=m^2)  \equiv g \sqrt{N_1}$
where $N_1$ is the normalization of the one-body Fock component $\phi_1$.

\section{The anomalous magnetic moment in the Yukawa model}
\begin{figure}[btph]
\begin{center}
\includegraphics[width=25pc]{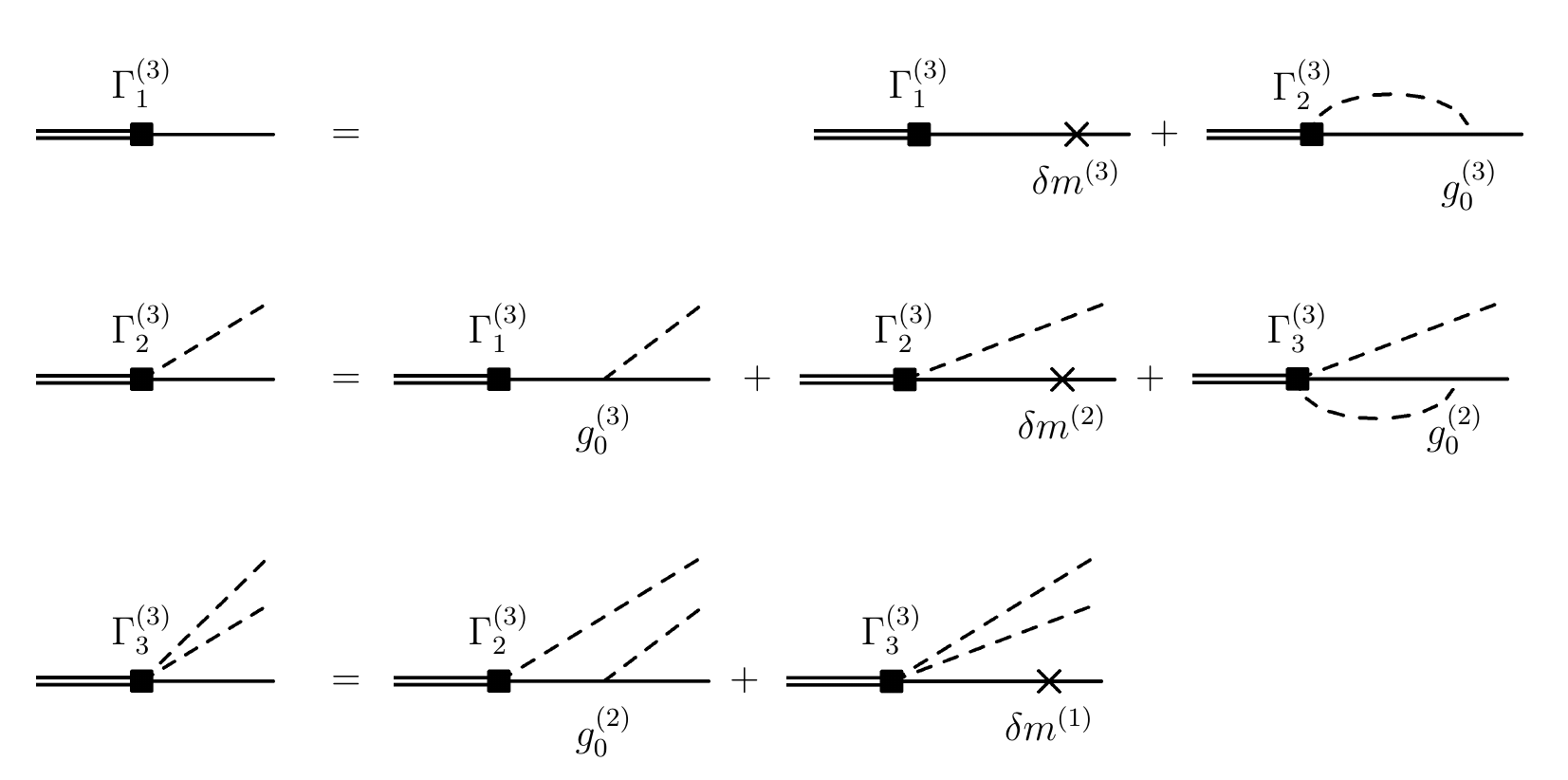}
\caption{System of equations for the
vertex functions
in the Yukawa model for the three-body Fock space truncation.
Dashed lines correspond to scalar bosons. \label{scalarfig}}
\end{center}
\end{figure}

In order to address the calculation of a true non-perturbative
system, we investigate the system composed of a fermion coupled to
scalar bosons for the three-body,
$N=3$, Fock space truncation \cite{kms_09}. The system of equations one has to solve is given in Fig.
\ref{scalarfig}, where the use of Fock sector dependent counterterms is shown. 
We use the Pauli-Villars (PV) regularization scheme, as detailed in \cite{kms_08}. The anomalous magnetic moment is calculated as a function of the boson PV mass, in the limit of infinite PV fermion mass.
\begin{figure}[btph]
\begin{center}
\includegraphics[width=18.5pc]{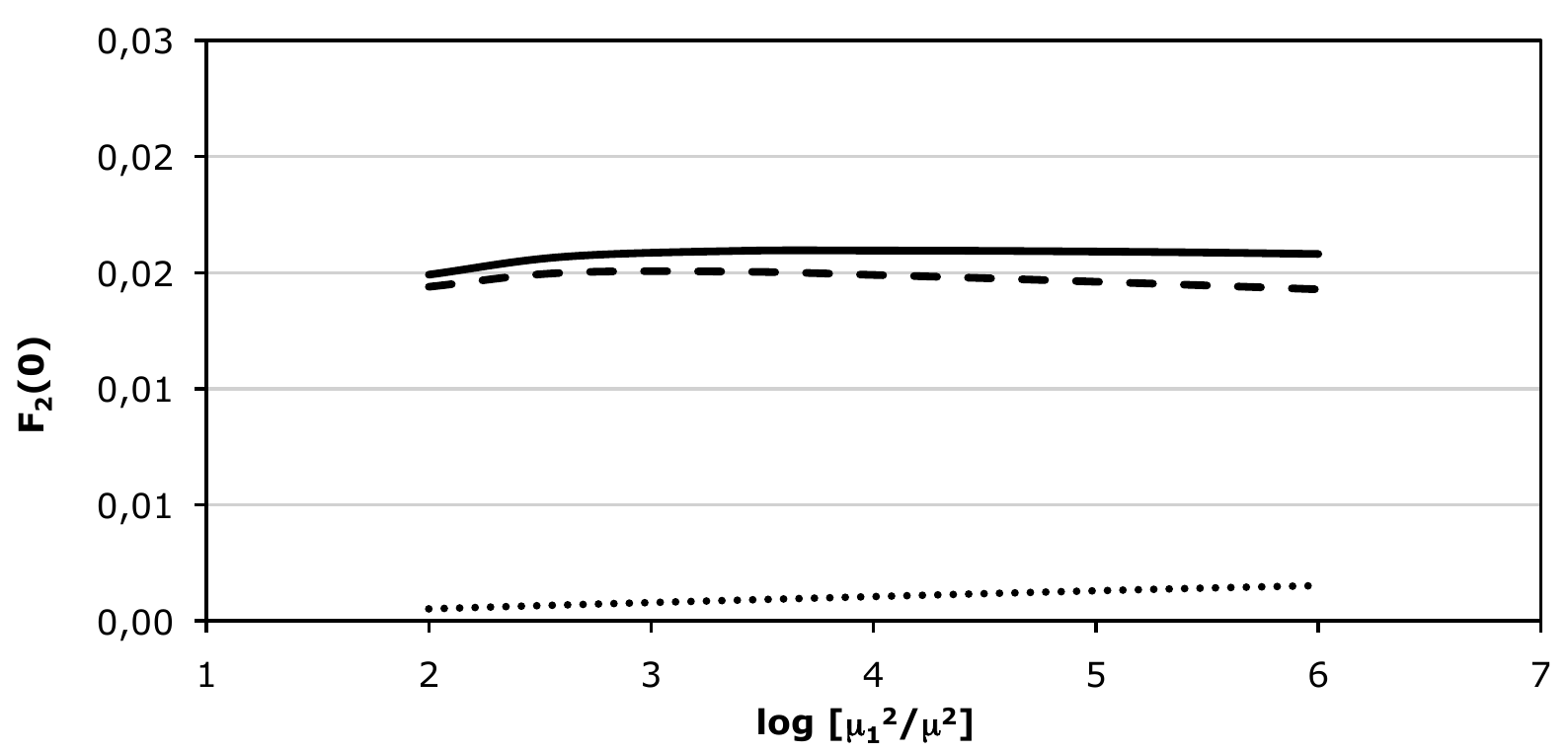} 
\includegraphics[width=17pc]{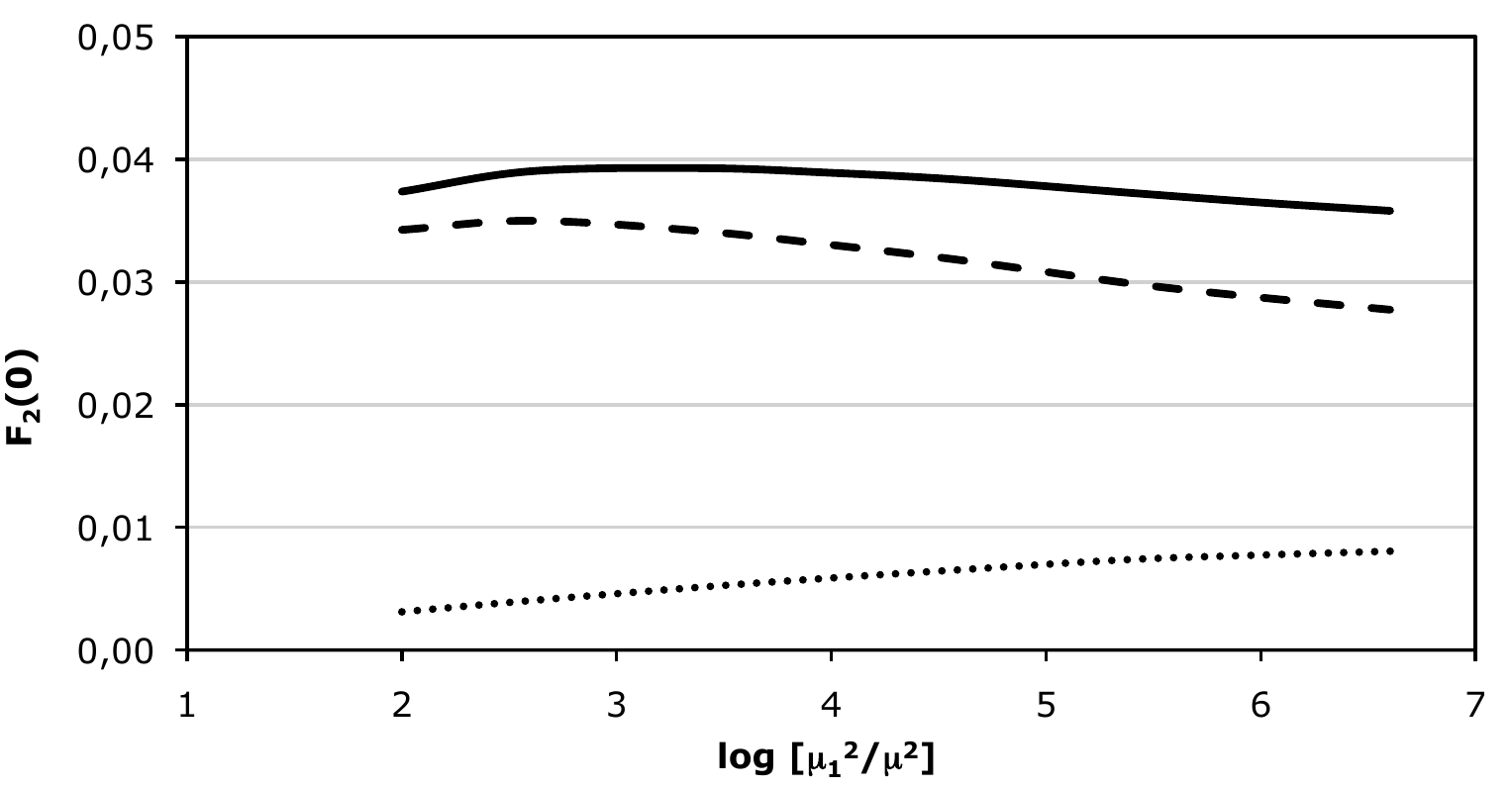} 
\caption{The fermion anomalous magnetic moment as a function of
the Pauli-Villars boson mass  $\mu_1$
for the $N=3$  Fock space  truncation. We
separate the contributions from the two-body (dashed line) and
three-body (dotted line) vertex function to the total result
(solid line). The lines are just drawn to guide the eyes. The results on the left correspond to $\alpha=0.2$ while the results on the right correspond to $\alpha=0.5$. The calculation are done with a fermion ($m$) and a boson ($\mu$) mass of $1$ GeV. \label{anomalous}}
\end{center}
\end{figure}
While the calculation for a boson-fermion coupling constant $\alpha=0.2$ shows nice convergence as a function of the PV mass, in the limit of large masses, the results for $\alpha=0.5$ are an indication that higher order Fock components may start to be sizeable.

\section{Perspectives}
It has been proposed recently \cite{OPVD,GW} to use the definition of quantum fields as operator valued distributions acting on test functions as a regularization/renormalization procedure. This construction is particularly suited for light-front calculations since the test functions can be included from the very begining in the definition of the Fock components, leading to a very transparent, and general, formulation \cite{MG}. Let us outline here the main steps of this formulation.
The physical field $\varphi(x)$ is defined in terms of the translation, $T_x$,  of the original distribution $\Phi$ acting on the test function $\rho$:
\begin{equation}
\varphi(x) \equiv T_x \Phi(\rho) =\int d^4y \phi(y) \rho(x-y)\ .
\end{equation}
We shall concentrate here on distributions singular in the UV domain.  Any physical amplitude can be written in a schematic way (in one dimension for simplicity) as:
\begin{equation} \label{amlitude}
{\cal A}=\int_0^\infty dX\ T(X)\  f(X)\ ,
\end{equation}
where $f$ is the Fourier transform of the test function, and $T(X)$ is a distribution which may lead, without any regularization procedure, to a divergent integral in the UV domain when $X \to \infty$. In LFD, $X$ is for instance proportional to ${\bf k}^2$, where ${\bf k}$ is the three-momentum of any of the constituents.
The choice of the test function - of compact support and with all its derivatives equal to zero at the boundary - should be done with care. In particular, one should make sure that the physical amplitudes are independent of the choice of the test function. This is achieved by using test functions which are partitions of unity, i.e. functions which are $1$ everywhere except at the boundary where they should go to zero \cite{GW}. The choice of the boundary of the test function, $H$, should also respect scaling invariance which tells us  that the limit $X\to \infty$ can be reached in many different ways since in this limit $\eta^2 X\to \infty$ for an arbitrary scale $\eta^2 >1$. In order to do that, it is necessary to consider a boundary condition for which $H$ depends on the kinematical variable $X$ \cite{GW}. We can choose, for example:
\begin{equation}
H(X)\equiv \eta^2X^\alpha\ . 
\end{equation}
In the limit where $\alpha \to 1^-$, the maximum value of X, defined by $H(X_{max})=X_{max}$ goes to infinity, and the test function goes to $1$ in the whole kinematical domain.
The extension of the distribution $T(X)$ in the UV domain can thus be done easily using the general Lagrange formula given by
 \begin{equation} \label{la3}
f(X)=-\frac{X}{k!}\int_1^\infty \frac{dt}{t} (1-t)^k \partial_X^{k+1} \left[ X^k f(Xt) \right]\ ,
\end{equation}
for any integer $k \ge 0$. 
The physical amplitude  writes in that case:
\begin{equation}
{\cal A}=\int_0^\infty dX \ \frac{(-1)^k}{k!} X^k\partial_X^{k+1} \left[ X T(X)\int_1^{\eta^2 X^{\alpha-1}} \frac{dt}{t}(1-t)^k\right] f(Xt)  \to \int_0^\infty dX \ \tilde T(X)\ . \label{Ato}
\end{equation}
in the limit where $\alpha \to 1^-$. This defines the extension $\tilde T^>(X)$ of the distribution $T(X)$ in the UV domain. The value of $k$ to be used depends on the degree of singularity of the initial distribution. The amplitude (\ref{Ato}) is now completely finite. 
Note that we do not need the explicit form of the test function in the derivation of the extended distribution $\tilde T^>(X)$. We only rely on its mathematical properties and on its dynamical construction. 


\end{document}